\documentclass[11pt]{article}

\usepackage[margin=1in]{geometry}
\usepackage{microtype}

\usepackage{amsmath, amssymb, amsthm, amsfonts, mathtools}

\usepackage[T1]{fontenc}
\usepackage{lmodern}

\usepackage{listings}

\usepackage[colorlinks=true, linkcolor=blue, citecolor=blue, urlcolor=blue]{hyperref}

\theoremstyle{remark}

\usepackage{xcolor}
\definecolor{keywordcolor}{rgb}{0.1, 0.2, 0.6}    
\definecolor{tacticcolor}{rgb}{0.1, 0.2, 0.6}    
\definecolor{commentcolor}{rgb}{0.4, 0.4, 0.4}   
\definecolor{symbolcolor}{rgb}{0.0, 0.0, 0.0}
\definecolor{sortcolor}{rgb}{0.1, 0.2, 0.6}    
\definecolor{attributecolor}{rgb}{0.7, 0.1, 0.1} 

\usepackage{listings}

\usepackage{upgreek}
\usepackage{inconsolata}

\usepackage{todonotes}

\usepackage{enumitem}
\setlist{noitemsep}

\usepackage{authblk}

\title{Formal verification of the S-two AIR}

\author[1]{Jeremy Avigad}
\author[2]{Anat Ganor}
\author[2]{Lior Goldberg}
\author[2]{David Levit}
\author[2]{Ohad Nir}
\author[ \hspace{-1ex}]{Yoav Seginer}  
\author[2]{Alon Titelman}

\affil[1]{Carnegie Mellon University}

\affil[2]{StarkWare Industries Ltd.}

\date{\today}

\begin{document}
\maketitle

\begin{abstract}
StarkWare's S-two prover provides an efficient means for establishing, on blockchain, that a program written in the Cairo virtual machine language runs to completion. The latter claim is encoded by an algebraic intermediate representation (AIR) that captures the semantics of the Cairo language. The AIR asserts the existence of tables of values from a finite field satisfying certain algebraic constraints. A cryptographic interactive proof system, circle STARK \cite{habock:levit:papini:24}, provides an efficiently-checked certificate that the AIR is satisfied. We describe our verification, using the Lean 4 proof assistant, that the AIR encoding is sound, which is to say, the satisfiability of the AIR implies the computational claim.
\end{abstract}

\section{Introduction}

StarkWare's Cairo ecosystem provides an efficient means of verifying smart contract execution on blockchain. A prover and a verifier begin by agreeing on a Cairo program whose successful termination establishes a desired claim. The program is compiled to instructions for the Cairo virtual machine, and the claim that such a program runs to completion is encoded by an algebraic intermediate representation (AIR) that captures the semantics of the Cairo language. We refer to this AIR as the Cairo-AIR. The Cairo-AIR asserts the existence of tables of values from a finite field satisfying certain algebraic constraints, some of which are specific to the Cairo program and inputs of interest. A cryptographic proof system provides an efficiently-checked certificate that the prover has a satisfying instance in hand, and this certificate is published and verified on blockchain.

Trust in these certificates requires trust in every step of the process: that successful termination of the original Cairo program implies the desired claim; that the compilation to machine code preserves the intended semantics of the Cairo program; that satisfiability of the Cairo-AIR implies that the machine code terminates successfully; and that the STARK certificate warrants the satisfiability of the Cairo-AIR constraints.

StarkWare uses a battery of techniques to ensure that these claims hold, including static testing, code review, regular audits, and peer review of the cryptographic protocols. This paper is about yet another method, namely, formal verification using the Lean 4 proof assistant \cite{de:moura:ullrich:21}. An interactive proof assistant allows users to write arbitrary mathematical definitions, theorems, and proofs in a formal language, and to establish correctness with respect to a formal axiomatic system. Users work interactively with a proof assistant to write proofs that are machine-checked for correctness. Assistance provided by symbolic and neural AI is steadily improving, though the process of writing formal proofs remains a significant undertaking. On the positive side, Lean's language is rich enough to express any conventional mathematical claim, its axiomatic system is strong enough to model conventional mathematical reasoning, users can draw on a large library, Mathlib \cite{mathlib}, of formalized mathematics, and Lean's trusted computing base provides a strong standard of correctness.

To date, we have used Lean to verify the soundness of the original AIR encoding of the Cairo virtual machine semantics, used by the \emph{Stone} prover \cite{avigad:et:al:22,goldberg:et:al:21}. We have provided end-to-end verification of several library functions, including dictionary lookups and elliptic curve signature validation written in CairoZero (a predecessor to the current version of Cairo), that are still in use \cite{avigad:et:al:25}. Using similar methods, we have verified several library functions, or \emph{Libfuncs}, for the current version of Cairo \cite{starkware:formal:proofs}.

Here we describe another verification project, namely, the verification of the much more elaborate and efficient AIR encoding used by StarkWare's new \emph{S-two} prover. The AIR encoding is based on an architecture with several components that call, and are called by, other components. We have verified not only that the algebraic constraints in individual components guarantee their specifications, but also that the algebraic encoding of the communication protocol between components is sound, and that, taken together, these specifications ensure the existence of an execution trace that shows that the Cairo machine code runs to completion.

In Section~\ref{section:background}, we provide background on proof assistants, the Cairo virtual machine, and the types of guarantees provided by the cryptographic protocol associated with S-two. In Section~\ref{section:soundness}, we explain the formal statement of our soundness theorem, and in Section~\ref{section:proof}, we describe some of the more interesting parts of the proof. Section~\ref{section:logups} breaks out an important ingredient, the logup protocol, which provides a general scheme for sharing information between components in an efficient way. The formal proofs are online at \url{https://github.com/starkware-libs/formal-proofs}.


\section{Background}
\label{section:background}

\subsection{Interactive proof assistants and formal verification}

Since the early twentieth century, it has been understood that mathematical arguments can be represented in formal axiomatic systems, allowing for proofs whose correctness can, in principle, be checked mechanically. Interactive proof assistants make this possible in practice, allowing users to write definitions and proofs in a proof language, like a programming language. The steps of the proof are checked against the axioms and rules of the underlying foundational proof system by carefully-written code that is known as the \emph{kernel}, or \emph{trusted computing base}. Proof assistants are elaborate pieces of software, including infrastructure and automation that help users construct complex proofs, but these do not need to be trusted; if faulty automation produces an invalid proof, that proof will be rejected by the kernel. The availability of independently-written reference checkers provides additional confidence that a formal claim has a valid proof.

For this project, we used the Lean 4 programming language and proof assistant \cite{de:moura:ullrich:21}. Lean is based on a version of dependent type theory whose logical strength is just slightly stronger than Zermelo--Fraenkel set theory, which mathematicians commonly view as the ``official'' foundation for mathematics. A typed foundation, like a typed programming language, has several benefits, often providing better error-checking and more compact forms of expression \cite{avigad:26}. The project was further supported by Lean's communally developed and maintained mathematical library, Mathlib \cite{mathlib}.

A proof assistant provides a very strong guarantee that a formally-specified claim has a formal proof according to the axioms and rules of the axiomatic foundation. The result is only valuable insofar as the formally-specified claim has the intended interpretation. For example, if the definition of integer multiplication in the proof assistant doesn't represent integer multiplication, the theorem \lstinline{∀ x y, mul x y = mul y x} doesn't establish the commutativity of integer multiplication. Confidence in the correctness of definitions in the underlying library is bolstered by showing that they have all the expected properties.

The main goal of this paper is to provide an overview of the soundness theorem we have verified in Lean, as well as the definitions and underlying theory that are needed to state and prove that theorem.

\subsection{The Cairo virtual machine}

The Cairo machine model is based on a simple CPU architecture with
three registers: a \emph{program counter} (pc), which points to the current
instruction in memory, a \emph{frame pointer} (fp), which generally points to the location of the local variables in a function call, and an \emph{allocation pointer} (ap), which generally points to the next free value in global working memory. A machine instruction consists of three 16-bit words, which generally serve as memory offsets for operations performed on memory, and 15 1-bit flags, which determine the nature of the instruction. The architecture is described in detail in the Cairo whitepaper \cite{goldberg:et:al:21}, and we have formalized the machine model and its semantics in Lean \cite{avigad:et:al:22}.

A notable feature of the model is that elements of memory, as well as the contents of the registers, are elements of the field of integers modulo a certain prime number, namely, $2^{251} + 17 \cdot 2^{192} + 1$. The CPU can therefore add and multiply values, but it cannot ask whether one value is greater than another. Cryptographic primitives described below can be used to assert that the contents of a memory location represents the cast of an integer in a certain range. In StarkWare terminology, ``casm'' is short for ``Cairo assembly language,'' which provides convenient representations of virtual machine code instructions. Thus, below and in our formalization, a ``casm state'' refers to a triple $(\mathit{pc}, \mathit{fp}, \mathit{ap})$ of elements of the underlying field.

Another notable feature of the machine model is that the memory is read-only.
To establish a computational claim on blockchain, the S-two prover makes public a partial assignment to memory that typically includes the program that is executed and the agreed-upon input and output. The prover also makes public the initial and final state of the registers and a bound on the number of steps, and produces a certificate that is intended to establish that there is an assignment of values to memory extending the partial one and an execution trace from the start state to the final state, according to the virtual machine semantics. Our goal here is to verify the soundness of the Cairo-AIR encoding used to specify this certificate.

\subsection{Components and messages}
\label{subsection:components:and:messages}

As explained by the S-two whitepaper \cite{habock:levit:papini:24}, the S-two AIR is best understood in terms of \emph{components} exchanging \emph{messages} over a message channel. The S-two AIR works over the field $\mathsf{M31}$ -- the field of integers modulo the prime $2^{31} - 1$, a value chosen carefully for efficient cryptographic computation. Each component $C$ is represented by a collection of polynomial constraints over \emph{witness variables} $X_1, \ldots, X_{M_C}$ in the field. Two tuples of polynomials $U_1(\vec X), \ldots, U_{R_C}(\vec X)$ and $V_1(\vec X), \ldots, V_{S_C}(\vec X)$ over the witness variables are also associated with each component; the former are called the \emph{inputs}, or \emph{uses}, of the component, and the latter are called the \emph{outputs}, or \emph{yields}. These polynomials are often as simple as a single variable, which is to say, it is often the case that certain witness variables are designated as inputs and others as outputs. The existence of values $X_1, \ldots, X_{M_C}$ satisfying the polynomial constraints expresses a relationship between the inputs and outputs. For example, a component with constraints $X_2 = X_1^2$ and $X_3 = X_2^2$, with $U_1(\vec X) = X_1$ and $V_1(\vec X) = X_3$, computes the input to the fourth power.

The S-two cryptographic protocol establishes the existence of tables of values for each component $C$, in other words, a table consisting of $N_C$ rows of tuples, each satisfying the corresponding polynomial constraints. You can think of each row as corresponding to a procedure call in the verification of the global claim established by the AIR. The values are elements of $\mathsf{M31}$. As we will explain, the S-two AIR encoding is designed so that the existence of tables of values satisfying the polynomial constraints and the use-yield correspondences implies the existence of an execution trace of the Cairo machine code.

What enables communication between components is that every use of one component is a yield of another; more precisely, every use tuple in the message channel must have a corresponding yield tuple with the same values. We refer to a tuple of use values specified by a component as a \emph{lookup}. Some components have only yield values, and are referred to in the S-two whitepaper as \emph{lookup components}, since their function is to provide values for other components to use. For example, a range-check component is a lookup component that simply yields the values $0, \ldots, M$ for some fixed $M$. Moreover, the protocol allows us to specify \emph{external lookups} -- a sequence of yield tuples $\vec I_1, \ldots, \vec I_k$, and another sequence of use tuples $\vec O_1, \ldots, \vec O_\ell$. These external lookups are added to the message channel alongside those contributed by the components, and are subject to the same use-yield matching.

We will see in Section~\ref{subsection:components} that a collection of components in the S-two AIR encoding are designed to certify pairs $(s, s')$ of registers where $s'$ is a successor state to $s$ in the Cairo virtual machine semantics. The ultimate goal of the encoding is to establish the existence of a sequence of states $s_0, s_1, \ldots, s_n$ such that $s_0$ is the initial state, $s_n$ is the final state, and for each $i < n$, the pair $(s_i, s_{i+1})$ is certified by one of these components. The underlying cryptographic protocol has a special variant of the message-passing protocol, chain lookups, that ensures that the pairs $(s, s')$ verified by the relevant components chain together in this way. This is achieved by requiring, for these lookups, that the multiset of use values \emph{equals} the multiset of yield values, rather than the weaker subset property used elsewhere, and by introducing the initial state $s_0$ and the final state $s_n$ as external lookups.

Finally, the AIR encoding includes lookup components with fixed values that the prover and verifier agree to, using a mechanism known as \emph{preprocessed columns} \cite{habock:levit:papini:24}. For example, a \lstinline{RangeCheck} component yields all values in the range $[0, M)$ for a fixed natural number $M$. This allows any other component to check whether a variable is in the range by a \emph{use} lookup.

In sum, the S-two AIR encoding consists of a finite list of components, each with associated constraint polynomials, designated use values, and designated yield values \cite[Definition 9]{habock:levit:papini:24}. An instance of the AIR is a collection of tables of values in $\mathsf{M31}$, one for each component. Such an instance is said to \emph{satisfy} the AIR (cf.~\cite[Definition 10]{habock:levit:papini:24}), if the following hold:
\begin{itemize}
  \item Each row of each component table satisfies the corresponding polynomial constraints.
  \item For each use tuple there exists an identical yield tuple (in any of the components or the external lookups).
  \item For the chain lookups, the multiset of use values equals the multiset of yield values.
  \item The preprocessed columns contain the agreed-upon values.
\end{itemize}

To a first approximation, our soundness theorem says that if a given instance of the S-two AIR encoding satisfies the Cairo-AIR with respect to a partial assignment to memory, then there is a full assignment to the memory and an execution trace of the Cairo virtual machine from the start state to the final state, where each successive state is determined by the instruction in memory according to the Cairo semantics. The actual proof is more delicate, because the cryptographic protocol used for establishing the lookup claims is a probabilistic argument based on a subtle algebraic computation \cite{habock:22}.
In particular, it involves picking a pseudo-random field element from an extension field.
We show that if either of the lookup claims fails, there is only a small set of ``random'' values that would fool the protocol into acceptance.
This translates to a negligible probability of error (called the soundness error).
Thus, the conclusion of the soundness theorem only assumes that,
\begin{itemize}
  \item each component's polynomial constraints hold of the associated table,
  \item the logup polynomials are satisfied with respect to the relevant uses and yields, and
  \item the chosen ``random'' value lies outside the small bad set.
\end{itemize}
We describe the logup protocol and our verification thereof in Section~\ref{section:logups}.

\subsection{Methodology}
\label{subsection:methodology}

In the Stone AIR encoding, the polynomial constraints were given by a list of polynomials enumerated explicitly in the whitepaper \cite{goldberg:et:al:21}. The list was small enough that we could export them from Stone to Lean and reason about them manually in Lean.

The S-two AIR encoding is much more elaborate. To start with, it uses constraints over the field of elements modulo the prime $2^{31}-1$ to establish claims about computations over the field $2^{251} + 17 \cdot 2^{192} + 1$. In our verification and below we refer to elements of the first field as \emph{Felt}s, short for ``field elements,'' and elements of the second field as \emph{Felt252}s. Reasoning about the latter in terms of the former poses challenges. Moreover, the S-two prover achieves dramatic efficiency gains using the more elaborate encoding based on components and messages described in Section~\ref{subsection:components:and:messages}. Optimizations include custom handling of individual instructions, as well as more efficient memory lookups.

The S-two AIR encoding is built using an infrastructure for generating AIR constraints in a modular, composable way. Rather than write down polynomial constructs and declare uses and yields manually, one uses a library of code to generate this data. A central data structure, the \emph{AIR builder}, maintains the list of components, and the polynomial constraints and messages associated to each. Individual functions are called to add constraints and messages. These are reusable and composable: if a function implements a useful gadget, other functions can call it to specify the same behavior in other components.

Our goal, then, was to prove the soundness of the AIR encoding generated by the infrastructure. A strategy one might consider is to run the code, generate the polynomial constraints and messages, translate them to Lean, and reason about them there. That was not feasible, however. Not only is the encoding too large, but the raw list of polynomials lacks sufficient structure. In other words, it is impossible to reason about the correctness of the constraints without reasoning about the code that generates them: the \emph{reason} the constraints guarantee the existence of computation traces is that the code was written for that purpose, and our soundness proof had to reflect that reasoning.

The strategy we adopted, therefore, was to translate the relevant parts of the infrastructure code to Lean and prove that the constraints and messages generated by the Lean code have the right properties. Lean is a pure functional programming language, which means that procedures cannot modify any global state. This makes it easier to reason about Lean code, though it requires some extra effort to model stateful computation, as explained in Section~\ref{subsection:modeling:components}. We therefore had to be strategic in choosing which aspects of the code to model in Lean and which aspects to trust. Specifically:
\begin{itemize}
  \item We modeled the constraints, yields, and lookups generated by each component in Lean.
  \item We verified the logup protocol in Lean.
  \item We did not model global aspects of the infrastructure, such as the general mechanisms that collect all the lookups from all the various components in one place, or the code that arranges the lookups into constraints. The assumption that these aspects of the code are correct is reflected in the hypotheses of our soundness theorem, as described in Section~\ref{section:soundness}.
\end{itemize}
This allowed us to focus on those aspects of the AIR generation where subtle errors were most likely to occur.

\section{The soundness theorem}
\label{section:soundness}

\subsection{The Cairo virtual machine semantics}

In Lean, the notion of a register state over an arbitrary field \lstinline{F} is defined as follows:
\begin{lstlisting}
structure RegisterState (F : Type) where
  pc : F
  ap : F
  fp : F
\end{lstlisting}
In the Cairo virtual machine, \lstinline{F} is the field of integers modulo the Felt252 prime, defined as follows:
\begin{lstlisting}
def Felt252Prime : Nat := 2^251 + 17 * 2^192 + 1
def Felt252 := ZMod Felt252Prime
\end{lstlisting}
Here \lstinline{ZMod n} is the Mathlib definition of the integers modulo \lstinline{n}. In our verification of the AIR encoding of the Stone prover \cite{avigad:et:al:22}, we defined the semantics of the AIR encoding, culminating in a definition of the next state relation for the virtual machine, over an arbitrary finite field \lstinline{F}:
\begin{lstlisting}
  def NextState (mem : F → F) (s t : RegisterState F) : Prop := ...
\end{lstlisting}
This definition expresses the semantics defined informally in the Cairo whitepaper \cite{goldberg:et:al:21}. Note that memory is modeled as a function from an arbitrary address in \lstinline{F} to a value in \lstinline{F}. The return type, \lstinline{Prop}, indicates that \lstinline{NextState mem s t} is a proposition, which is to say, an assertion that can be true or false. We use exactly the same specification in stating the soundness of the AIR encoding for S-two. In other words, the semantics of the virtual machine has not changed in the passage from Stone to S-two.

If the ultimate goal is to verify the soundness of S-two with respect to the Cairo semantics, we need to trust that the formal definition \lstinline{NextState} correctly defines that semantics. In previous work \cite{avigad:et:al:25}, we verified that certain instances of compiled Cairo machine code meet high level specifications with respect to the Cairo virtual machine semantics, and hence with respect to the AIR. In applications like that, it does not matter what virtual machine semantics a reader may have in mind, since the formal semantics is only a stepping stone, linking the final AIR encoding to the claim it is meant to establish.

\subsection{Expressions and their semantics}
\label{subsection:expression:and:their:semantics}

Recall that we use the term \emph{Felt} for elements of the field modulo the M31 prime, and the term \emph{Felt252} for elements of the field modulo the 252-bit prime used in the specification of the Cairo virtual machine. The STARK certificate establishes the existence of tables of Felts, one table for each component of the encoding. Each row of the table is a tuple of Felts, and with respect to each row, some \emph{use} values are specified as tuples of multivariate polynomials of elements of the row. The component's \emph{yield} values are specified in the same way.

The specification of a component is therefore a specification of the polynomial constraints that the elements are intended to satisfy, as well as the tuples of polynomial expressions that determine the uses and yields. The corresponding infrastructure code, and the Lean code we use to model it, generates the relevant polynomial expressions. The S-two \emph{prover} computes values satisfying the polynomial constraints and the use-yield correspondences, and generates a cryptographic certificate that they meet this specification. The S-two \emph{verifier} uses the cryptographic protocol to verify that the prover has done what it is supposed to do.

Our soundness theorem states, roughly, that the existence of tables of values satisfying the polynomial and message-passing constraints implies the existence of a Cairo execution trace consistent with a start state, end state, and partial assignment to memory that has been agreed upon by the prover and verifier. To state this formally, we need to say what a multivariate polynomial expression is, and what it means for a tuple of values to satisfy it. We express this in Lean as follows:
\begin{lstlisting}
inductive FeltExpr where
  | const (value : Felt) : FeltExpr
  | var (index : VarIndex) : FeltExpr
  | binary (op : BinaryOp) (left : FeltExpr) (right : FeltExpr) : FeltExpr
  | unary (op : UnaryOp) (child : FeltExpr) : FeltExpr
\end{lstlisting}
In other words, a \lstinline{FeltExpr} is either a constant Felt value, a variable $v_i$, or a compound expression consisting of a unary or binary operator applied to \lstinline{FeltExpr}. We also define what it means to \emph{evaluate} a \lstinline{FeltExpr} with respect to an assignment of Felts to the variables:
\begin{lstlisting}
def VarAssign := VarIndex → Felt

def FeltExpr.eval (varAssign : VarAssign) : FeltExpr → Felt
  | const value => value
  | var index => varAssign index
  | binary op left right =>
    let l := eval varAssign left
    let r := eval varAssign right
    match op with
    | BinaryOp.add => l + r
    ...
  ...
\end{lstlisting}
A variable assignment is a map from variable indices to values. A constant evaluates to itself; to evaluate a variable, one looks up the assignment; one evaluates a compound expression by evaluating the components and applying the relevant operation.

We note that this evaluation function is used to express the \emph{semantics} of the code we verified, and is not part of the code itself. This is a common pattern in our formalization: first, we modeled the relevant S-two data structures and procedures in Lean, and then we expressed the intended semantics. In particular, a polynomial constraint is expressed as a \lstinline{FeltExpr}, \lstinline{c}, and the constraint is satisfied with respect to a variable assignment \lstinline{v} if \lstinline{c.eval v = 0}.

The statement of the soundness theorem requires nothing more than the semantic relation we have just described, namely, the notion of an assignment satisfying a polynomial expression. The \emph{proof} of the theorem, on the other hand, requires porting several additional data structures from the infrastructure code into Lean, some of which we describe in Section~\ref{section:proof}.

\subsection{Components and their semantics}
\label{subsection:components}

The infrastructure for constructing AIRs allows code to add components dynamically. However, modeling these mechanisms in Lean and reasoning about them formally is complicated and orthogonal to the soundness concerns we are trying to address. Thus our Lean version of the code has a fixed set of components:
\begin{lstlisting}
inductive Component where
  | Opcode (o : Opcode) : Component
  | RangeCheck : Component
  | MemoryAddrToId : Component
  | MemoryIdToValue : Component
  | VerifyInstr : Component
\end{lstlisting}

We now give a short overview of the different components.

The components \lstinline{RangeCheck}, \lstinline{MemoryAddrToId}, and \lstinline{MemoryIdToValue} are \emph{lookup} components, which is to say, they add only yields to the message buffer and have no use lookups.
They all use the \emph{preprocessed columns} primitive, which allows the prover and verifier to agree on fixed values for some of the component variables.

\lstinline{RangeCheck}: The Lean implementation of the \lstinline{RangeCheck} component yields assertions to the effect that a pair $(x, k)$, where $x$ is a Felt and $k$ is a natural number, satisfies $0 \le x < 2^k$. The actual S-two implementation has several specialized range-check components, such as \lstinline{rc20} for 20-bit values and \lstinline{rc9_9} for pairs of 9-bit values (an efficiency optimization). In our Lean modeling we collapse these into a single component, as described here.

\lstinline{MemoryAddrToId} and \lstinline{MemoryIdToValue}: \lstinline{MemoryAddrToId} registers pairs $(x, i)$ mapping a Felt memory address $x$ to a Felt id $i$, and \lstinline{MemoryIdToValue} registers pairs $(i, y)$ mapping a Felt id $i$ to a Felt encoding of a Felt252 value $y$. The representation of each Felt252 is checked for validity (cf.~Section~\ref{subsection:encoding:felt252s}). The use of an intermediate id enables an optimized component for copying a value from one memory address to another, a pattern that occurs frequently when passing arguments to functions. The infrastructure code is designed so that the mapping from $x$ to $i$ and the mapping from $i$ to $y$ are both well-defined functions. This is enforced by using a preprocessed column whose values (an address for \lstinline{MemoryAddrToId}, or an id for \lstinline{MemoryIdToValue}) are distinct in every row of the table. We do not verify these aspects of the infrastructure code; our soundness theorem assumes these properties as hypotheses, and the contents of these preprocessed columns must be verified separately.

\lstinline{VerifyInstruction}: Viewed as a procedure, the \lstinline{VerifyInstruction} component takes a tuple
$(c, o_1, o_2, o_3, f_1, f_2, \ldots, f_{15})$ consisting
of 19 Felts and adds constraints that ensure that the Felt252 at location $c$ in memory encodes an instruction with 16-bit offsets $o_1$, $o_2$, and $o_3$ and boolean flags $f_1, \ldots, f_{15}$. The component issues use lookups to the memory and range-check components as needed. It therefore uses values in the message buffer tagged as coming from these components, and yields tuples consisting of 19 Felts satisfying the corresponding constraints.

\lstinline{Opcode} components: each of the opcode components takes as input a Cairo register state $(pc, ap, fp)$, and returns an output state $(pc', ap', fp')$ satisfying the constraints that $(pc', ap', fp')$ is the next register state after $(pc, ap, fp)$ in a Cairo execution trace, for a particular opcode. In other words, each of them uses the memory, \lstinline{RangeCheck}, and \lstinline{VerifyInstruction} components, checks that the opcode is as expected, and adds constraints guaranteeing that the output register state stands in the proper relation to the input one.
Chain lookups, described in Section~\ref{subsection:components:and:messages}, use a special mode in which the multiset of use values is required to equal the multiset of yield values, rather than the usual subset relation.
Using a chain lookup, each component \emph{uses} a 3-tuple $(pc, ap, fp)$ for the input state and \emph{yields} a 3-tuple $(pc', ap', fp')$ for the output state, ensuring that these uses and yields chain together to form a correct execution trace.

Each component has an associated set of constraints and lookups. To each component, we associate an \lstinline{AirBuilder} structure, which contains the constraints that are generated by the infrastructure code, and an \lstinline{AirLookupTerms} structure, which includes both the use terms read from the message buffer and the yield terms added to the buffer. We collect them all in the following function:
\begin{lstlisting}
def ComponentLookupCall (c : Component) : AirBuilder × AirLookupTerms := ...
\end{lstlisting}
If \lstinline{ab} is an \lstinline{AirBuilder}, its list of constraints, \lstinline{ab.constraints}, is given by an array of Felt expressions, and the AIR builder is satisfied with respect to a variable assignment if all these Felt expressions evaluate to 0:
\begin{lstlisting}
def AirBuilder.SatisfiedBy (ab : AirBuilder) (varAssign : VarAssign) : Prop :=
   ∀ i, (h : i < ab.constraints.size) → ab.constraints[i].eval varAssign = 0
\end{lstlisting}
A table of rows for a component is represented as an array of variable assignments. Remember that the soundness theorem assumes that for each component, each row of the component's table satisfies the component's constraints. This is expressed as follows:
\begin{lstlisting}
def ComponentsSatisfied [Fact (Nat.Prime Stwo.P)]
    (varAssigns : Component → Array VarAssign) :=
  ∀ c, ∀ v ∈ varAssigns c, (ComponentLookupCall c).1.SatisfiedBy v
\end{lstlisting}
The argument \lstinline{[Fact (Nat.Prime Stwo.P)]} asserts that the M31 prime used by S-two is in fact a prime number, something we do not verify in Lean but carry along as a hypothesis. We do the same for the assumption \lstinline{[Fact (Nat.Prime Felt252Prime)]} about the Felt252 prime. These assumptions are simply dragged along to the final soundness theorem; below, we generally omit them for brevity.

\subsection{Messages and their semantics}
\label{subsection:messages}

Every component defines a set of tuple expressions which are the use and yield values of the component. The yield tuples determine the \emph{relation} that is defined by the component; for example, the yields of the \lstinline{VerifyInstruction} component are tuples that have the property of representing the instruction at the given memory location. In general, the uses and yields are tuples of multivariate polynomials over the variables of the component, that is, tuples of \lstinline{FeltExpr}. For each such tuple the component specifies whether it is a use or yield tuple and which relation (component) it belongs to.

\begin{lstlisting}
inductive UseOrYield where
  | use   : UseOrYield
  | yield : UseOrYield

structure LookupTerm {n_r : ℕ} (tuple_len_minus_one : Fin (n_r + 1) → Nat) where
  rel : Fin (n_r + 1) -- The relation id
  tuple : Fin (tuple_len_minus_one rel + 1) → FeltExpr
  useOrYield : UseOrYield
\end{lstlisting}
The function \lstinline{tuple_len_minus_one} specifies, for each relation, the length of all the tuples in that relation; we use \lstinline{length - 1} rather than \lstinline{length} to encode the fact that the length must be positive.
Every variable assignment to the component then induces an evaluation of the tuples of expressions (of that component) into tuples of \lstinline{Felt}s (in the way described in Section~\ref{subsection:expression:and:their:semantics}). By the nature of a communication system, which needs to establish relations between tuples generated by different components, we then put all the evaluations of all lookup terms of all components into one large list, where \lstinline{varAssigns} is a function which defines the array of variable assignments for each component.
\begin{lstlisting}
-- All evaluated lookups of all components.
def ComponentLookups (varAssigns : Component → Array VarAssign) := ...
\end{lstlisting}

The satisfaction by these tuples of the cumulative logup constraints, which are explained in Section~\ref{section:logups}, guarantees, with high probability, that certain relations hold between the use and yield tuples of each relation. This holds across all components. The cumulative logup constraints, as described in Section~\ref{subsection:polynomial:constraints}, are generated from the lookup tuple expressions, but we do not model this process explicitly. Instead, we show that the required properties hold (under appropriate assumptions) for any cumulative logup constraints. As long as these constraints encode the same tuples as those generated by the evaluation of the lookup terms of the components, the required relations hold for the tuples from the components.

The logup constraints are used to establish one of two properties of the use and yield tuples of a relation. For \emph{subset relations}, the logup constraints imply that the set of use tuples is a subset of the set of yield tuples. When one component establishes some property on the yield tuples, this implies that the same property also holds for the use tuples (for example, it is range checked or is a valid instruction). For \emph{chain relations}, the logup constraints establish that each tuple appears as often in the use tuples as it does in the yield tuples. Most importantly, this is used to establish that a sequence of valid state transitions results in a sequence of register states starting with the initial state and ending with the final state.

\subsection{Statement of the soundness theorem}
\label{subsection:soundness}
Our main soundness theorem for S-two, presented in full in Appendix~\ref{section:formal:statement}, looks very similar to our main soundness theorem for Stone \cite{avigad:et:al:22}. Roughly, it says that if the S-two AIR is satisfied with respect to some public data, there is a terminating Cairo execution trace consistent with that data. As with Stone, the public data shared by the prover and the verifier consists of the starting register state and ending register state, and a partial assignment to the memory. This is given by the parameter \lstinline{inp : InputData}, whose datatype is defined as follows:
\begin{lstlisting}
structure InputData where
  initialAp : Nat
  initialPc : Nat
  finalAp : Nat
  finalPc : Nat
  mStar : Felt252 → Option Felt252
\end{lstlisting}
Besides the Cairo-AIR satisfiability, the S-two verifier checks that the initial and final values of the frame pointer (fp) are both equal to the initial value of the allocation pointer (ap). This is reflected in the definitions of the initial and final state in the statement of the soundness theorem:
\begin{lstlisting}
def initialState (inp : InputData) : CasmStateVal :=
{ pc := inp.initialPc, ap := inp.initialAp, fp := inp.initialAp }

def finalState (inp : InputData) : CasmStateVal :=
{ pc := inp.finalPc, ap := inp.finalAp, fp := inp.initialAp }
\end{lstlisting}
Elements of the data type \lstinline{Option Felt252} are either of the form \lstinline{some x}, where \lstinline{x} is an element of \lstinline{Felt252}, or \lstinline{none}. Thus the field \lstinline{mStar} represents a partial assignment of Felt252 values to Felt252 memory addresses.

The S-two verifier also checks the partial assignment to the memory by introducing appropriate external lookups. The data to the soundness theorem thus includes \lstinline{pubMemLookups : PublicMem.Lookups}, where the latter data type is defined as follows:
\begin{lstlisting}
structure PublicMem.Lookups where
  lookups : Array (LookupTermVal rel_lengths)
  rel_in_lookup: RelInLookups lookups
  h_mem : AreMemoryUseLookups lookups
\end{lstlisting}
In other words, the public memory lookups are given by an array of lookups, each of which encodes the relation id in the tuple and each of which is a use value that corresponds to one of the memory components. The hypothesis
\begin{lstlisting}
h_pubMem : PublicMem.LookupsAgree inp.mStar pubMemLookups
\end{lstlisting}
states that these lookups agree with the partial memory assignment given by \lstinline{inp.mStar}, in the sense that every assignment has an entry in the lookups.

The next several assumptions concern the message-passing protocol. As explained in Section~\ref{section:logups}, we verify the correctness of the algebraic constraints and the protocol that ensures that all the uses are contained in the yields, with high probability. Generating the relevant constraints, however, requires certain bookkeeping in the infrastructure code that we do not verify, and hence, we leave the claim that the bookkeeping has been done correctly as an unverified assumption. Specifically, we need to assume:

\begin{lstlisting}
  -- Lookups are satisfied
  (h_satisfied : LookupsSatisfied t n_s NUM_LOOKUP_REL_MINUS_ONE rel_lengths chain_rels)
  -- Lookups agree with the assignments to the lookup terms
  (h_use_agree : ∀ rel, RelUseLookupsAgree inp pubMemLookups varAssigns rel (h_satisfied.tuples rel))
  (h_yield_agree : ∀ rel, RelYieldLookupsAgree inp pubMemLookups varAssigns rel (h_satisfied.tuples rel))
\end{lstlisting}
The first assumption is a structure that combines the logup constraints with the assumptions that these logup constraints are satisfied and that the random values used in these constraints do not fall into the (small) bad sets that need to be avoided. (See the discussion at the end of Section~\ref{subsection:components:and:messages} for details.) The next two assumptions (one for use tuples and one for yield tuples) tell us that the logup constraints described by the first assumption are constructed for the same use and yield tuples as those generated by evaluating the lookup terms of the components. In this way, the conclusions that follow from the satisfaction of the logup constraints described in the first assumption apply to the use and yield values generated for the components.

We now turn to assumptions about the memory and \lstinline{RangeCheck} lookup components. The S-two prover's job is to determine a consistent assignment to memory, extending the public partial assignment. Recall from Section~\ref{subsection:components} that this includes an intermediate assignment of Felt ids to Felt memory addresses. Soundness requires knowing that the map from addresses to ids and the map from ids to values are well-defined functions, which is to say, for every two pairs $(a, i)$ and $(a', i')$ registered as address-to-id lookups, if $a = a'$, then $i = i'$, and similarly for the id-to-value lookups. As explained in Section~\ref{subsection:components}, this is enforced by having the addresses (for \lstinline{MemoryAddrToId}) and the ids (for \lstinline{MemoryIdToValue}) come from a preprocessed column whose values are all distinct. We do not verify the correctness of this part, and we therefore add it as a hypothesis of the soundness theorem:
\begin{lstlisting}
  (h_is_mem_addr_to_id : IsMemAssign (varAssigns .MemoryAddrToId))
  (h_is_mem_id_to_value : IsMemAssign (varAssigns .MemoryIdToValue))
\end{lstlisting}
We also need to know that every entry in the range-check lookup component is checked to be in the interval $[0,2^n)$, for a specified value of $n$. This is expressed as follows:
\begin{lstlisting}
  (h_rc : IsRangeCheckAssign (varAssigns .RangeCheck))
\end{lstlisting}
Of course, we need to know that each component's tables satisfy the corresponding polynomial constraints,
\begin{lstlisting}
  (h_components: ComponentsSatisfied varAssigns)
\end{lstlisting}
using the \lstinline{ComponentsSatisfied} predicate defined in Section~\ref{subsection:components}.

Finally, to ensure that the registers do not overflow as Felt values (see Section~\ref{subsection:casm:address:bounds}), the next two assumptions bound the number of steps in the execution trace and the values of the registers in the initial state:
\begin{lstlisting}
  (h_num_steps :
    (RelLookups inp pubMemLookups varAssigns OPCODE_TRACE_REL_INDEX .use).size ≤ 2^29)
  (h_state_bound : (initialState inp).strongly_bounded₀ 0)
\end{lstlisting}
More precisely, rather than bound the number of steps directly, we bound the number of uses of the opcode lookup table. The second assumption amounts to the statement that the initial allocation pointer is less than $2^{29}-1$.

The conclusion of the soundness theorem is virtually the same as the conclusion of the soundness theorem for the Stone encoding \cite{avigad:et:al:22}. In words: there are an assignment to memory extending the partial input assignment and a sequence of register states of length at most $2^{29} + 1$, such that the first element of the sequence is the agreed-upon initial state, the last element of the sequence is the agreed-upon final state, and each successive state in the sequence satisfies the next state relation, with respect to the assignment to memory, as specified by the Cairo semantics.
\begin{lstlisting}
  ∃ mem : Felt252 → Felt252,
    Option.FnExtends mem inp.mStar ∧
    ∃ n : Nat, n ≤ 2^29
      ∧ ∃ exec : Fin (n + 1) → RegisterState Felt252,
          exec 0 = (initialState inp).toRegisterStateFelt252
          ∧ exec (Fin.last n) = (finalState inp).toRegisterStateFelt252
          ∧ ∀ i : Fin n, NextState mem (exec i.castSucc) (exec i.succ)
\end{lstlisting}
Putting it all together, we have the soundness theorem as stated in Appendix~\ref{section:formal:statement}.

There is a slight complication in the fact that the S-two prover uses Felts rather than Felt252s for the memory addresses and Cairo register values. (See the discussion in Section~\ref{subsection:casm:address:bounds}.) This is reflected in the cast \lstinline{toRegisterStateFelt252}, allowing us to interpret a Felt-based register state as a Felt252-based one. These restrictions mean that the S-two prover can only succeed in producing an AIR for Cairo programs for which the memory addresses and register states are small enough to be captured by a Felt. Similarly, our statement of the soundness theorem means that the prover can only succeed in producing an AIR for traces of length at most $2^{29}+1$. The statement that the prover can always succeed in producing an AIR under the circumstances is a property known as \emph{completeness}, which is dual to soundness. In other work, we have established similar completeness claims, but this work only addresses soundness: under the assumptions enumerated in this section, there exists a suitable execution trace.

\section{Proof of the soundness theorem}
\label{section:proof}

In this section, we provide impressionistic glimpses of some aspects of the proof of the soundness theorem, to convey a sense of the work involved.

\subsection{Encoding Felt252s}
\label{subsection:encoding:felt252s}

One of the most challenging aspects of the S-two AIR encoding is the fact that it uses constraints over the field of elements modulo the 31-bit Felt prime to establish claims about computations over the field of elements modulo the 252-bit Felt252 prime. StarkWare's engineers decided to represent each Felt252 value as a tuple of 28 Felts, each range-checked to be a 9-bit number, encoding 252 bits exactly. Thus we have the following:
\begin{lstlisting}
abbrev FELT252_N_WORDS := 28
def FELT252_BITS_PER_WORD := 9

def Felt252Expr := Fin FELT252_N_WORDS → FeltExpr
def Felt252Words := Fin FELT252_N_WORDS → Felt

def FeltExpr.eval (varAssign : VarAssign) : Felt252Expr → Felt252Words :=
  fun expr i => (expr i).eval varAssign
\end{lstlisting}
In other words, a \lstinline{Felt252Expr} is a tuple of 28 \lstinline{FeltExpr}s, and a \lstinline{Felt252Words} is a tuple of 28 Felts. Not every element of \lstinline{Felt252Words} is a valid encoding; we need to ensure that each Felt is less than $2^9$, or, more precisely, that each Felt is the cast of a natural number less than $2^9$ to the field of integers modulo the Felt prime. The expression \lstinline{x.IsRangeChecked xn} defined below asserts that \lstinline{x} is represented by such a tuple of natural numbers, \lstinline{xn}.

\begin{lstlisting}
def Felt252Nats := Fin FELT252_N_WORDS → Nat

def Felt252Words.IsRangeChecked (xn : Felt252Nats) (x : Felt252Words) : Prop :=
    ∀ i, x i = ↑(xn i) ∧ xn i < 2^9
\end{lstlisting}
Of course, the 28-tuple of natural numbers is intended to represent a single Felt252 value:
\begin{lstlisting}
def Felt252Nats.eval_nat (xn : Felt252Nats) : Nat :=
  ∑ i : Fin FELT252_N_WORDS, xn i * 2^(FELT252_BITS_PER_WORD * i)

def Felt252Nats.eval (xn : Felt252Nats) : Felt252 := eval_nat xn
\end{lstlisting}
The second definition uses an implicit cast from the corresponding natural number to the Felt252 data type.

This was a common pattern in our verification: AIR constraints are polynomials over the Felt field, but we generally want to think of calculations in the AIR as telling us something about the objects they are intended to represent. Consider an AIR component calculation over Felt values, and assume component values have been range-checked or otherwise guaranteed to represent integers in a certain range. Suppose an AIR calculation guarantees that Felts $x_1$ and $x_2$ represent natural numbers $n_1$ and $n_2$; we can show that $n_1 = n_2$ by showing $x_1 = x_2$ and verifying that $n_1$ and $n_2$ are both less than the Felt prime. More generally, we may have complex objects expressed in terms of Felts, but we want to reason about them at the level of natural-number representations. Once again, that is typically a matter of bounding calculations carefully and using equations in the Felt field to guarantee equations in the natural numbers. Such reasoning was ubiquitous in our verification (as well as in \cite{avigad:26}), and constituted much of our work.

Note that even with the representation of a Felt252 element by an element of \lstinline{Felt252Nats} in which each component is a 9-bit number, some Felt252s can be represented in more than one way. For example, $0$ and the Felt252 prime $2^{251} + 17 \cdot 2^{192} + 1$ are distinct 252-bit natural numbers, but evaluate to the same Felt252 value. StarkWare's engineers had to design component constraints to be sensitive to this fact, and our verification had to ensure that they had done so correctly.

\subsection{Modeling components}
\label{subsection:modeling:components}

StarkWare's infrastructure for building AIR encodings provides functions that one can call incrementally to construct components. A global data structure, the \emph{AIR builder}, maintains the list of components and the polynomial constraints and messages associated to each. The AIR builder is mutable: when a function is called to add constraints or messages, it modifies the AIR builder in place. Lean, on the other hand, is a pure functional programming language, which means that procedures cannot modify any global state. The fact that the value of a function on a set of arguments does not depend on any ambient state makes it easier to reason about Lean code, though it requires some extra effort to model stateful computation.

The conventional way to model stateful computation in a pure functional language is to make the state an explicit argument and return value of each procedure. Recall that the \lstinline{VerifyInstruction} component takes as input a casm address, three offsets, and 15 flags, and verifies that the offsets are in the range $0\ldots2^{16}-1$, the flags have value 0 or 1, and the instruction in memory at the given address corresponds to the given offsets and flags. In doing so, it issues lookups to the memory and range-check components as needed. The Lean code that generates this component has the following signature:





\begin{lstlisting}
def VerifyInstruction.call
    (airBuilder : AirBuilder)
    (lookupTerms : AirLookupTerms)
    (casmAddress : CasmAddress)
    (offset0 offset1 offset2 : FeltExpr)
    (flags : Fin 15 → FeltExpr) :
    AirBuilder × AirLookupTerms :=
    ...
\end{lstlisting}
Notice that the AIR builder and lookup terms are explicit arguments and return values. Functional programming aficionados will recognize the threading of state as an instance of the \emph{state monad}. As a programming language, Lean has built-in support for monadic programming, including notation that makes it possible to write such programs in conventional imperative style. However, when we were working on the project, Lean did not have good support for reasoning about programs written in this way, so we resorted to simulating monadic behavior by hand. Subroutines in the infrastructure code correspond to the composition of such functions in Lean, allowing us to prove our soundness theorems compositionally.

Note that the inputs to \lstinline{VerifyInstruction.call} are Felt \emph{expressions}, that is, symbolic expressions in variables that correspond to the AIR's columns. The desired specification is that for the corresponding Felt \emph{values}, under any assignment that satisfies the constraints and message-passing requirements, there is a valid instruction in memory at the given address, corresponding to the given offsets and flags. This specification of the Felt values is expressed as follows:

\begin{lstlisting}
def spec
    (memAssign : Felt252IdMemoryAssign)
    (casmAddress: CasmAddressVal)
    (offset0 offset1 offset2 : Felt)
    (flags : Fin 15 → Felt) : Prop :=
  ∃ instruction : Instruction,
    memAssign.HasInstruction casmAddress instruction ∧
    offset0 = instruction.offDst.toNat ∧
    offset1 = instruction.offOp0.toNat ∧
    offset2 = instruction.offOp1.toNat ∧
    flags 0 = instruction.dstReg.toFelt ∧
    flags 1 = instruction.op0Reg.toFelt ∧
    flags 2 = instruction.op1Imm.toFelt ∧
    ... ∧
    flags 13 = instruction.opcodeRet.toFelt ∧
    flags 14 = instruction.opcodeAssertEq.toFelt
\end{lstlisting}

The final constraints and lookups for each row of the AIR component table are generated by calling \lstinline{VerifyInstruction.call} starting with an empty AIR builder and lookup table. We prove a soundness theorem to the effect that for any variable assignment satisfying the constraints and any memory assignment consistent with the memory lookup calls, if the range-checked lookups satisfy the range-check constraints, then the evaluation of the casm address, offsets, and flags with respect to the variable assignment satisfy the specification above.

Recall that the soundness theorem described in Section~\ref{section:soundness} is stated in terms of the global message-passing protocol, which guarantees that if a component X depends on another component Y, then every time X uses a lookup of Y, the corresponding value is, in fact, a yield of Y. Thus, for example, the hypotheses on the range-check lookups and memory assignment lookups in the previous paragraph are satisfied because of global assumptions about the range-check and memory yields. In order to stitch our local soundness theorems together, we therefore reformulated each of them in terms of global assumptions and conclusions regarding the lookup tables. The relevant bookkeeping is somewhat verbose, but it is needed to pass from a local specification of each component's behavior to the global soundness claim, which depends on assumptions about all the components and messages at once.

This therefore summarizes our verification tasks:
\begin{enumerate}
  \item We modeled the AIR builder and lookup term data structures and infrastructure in Lean.
  \item We modeled the various S-two procedures used to generate AIR constraints in Lean, wrote specifications for each, and proved soundness. In particular, we obtained soundness proofs for each of the components described in Section~\ref{subsection:components}.
  \item We modeled and proved soundness for the message-passing protocol, ensuring that the yields of each component are sufficient to justify the uses of each component, with high probability.
  \item We put the pieces together to obtain the proof of the main soundness theorem.
\end{enumerate}

\subsection{Generic opcode}

Most opcode components in the Cairo-AIR encoding handle a specific opcode; this specialization reduces the number of table cells required for the verification of the corresponding step, by avoiding the need to decompose the machine instruction and interpret all the flags. The \lstinline{GenericOpcode} component is the catch-all: it covers all opcodes, at a higher cost. It takes a single casm state, looks up the corresponding instruction in memory, and calculates the next casm state. More precisely, it issues lookup calls to the range check component, the memory components, and the \lstinline{VerifyInstruction} component, and adds constraints that ensure that the output casm state bears the correct relationship to the input. The signature of the component call is therefore:

\begin{lstlisting}
def GenericOpcode.call
    (airBuilder : AirBuilder)
    (lookupTerms : AirLookupTerms)
    (casmState : CasmState) :
    AirBuilder × AirLookupTerms × CasmState := ...
\end{lstlisting}

The specification the component guarantees is relatively simple:
\begin{lstlisting}
def GenericOpcode.spec
    (memory : Felt252IdMemoryAssign)
    (casmStateVal : CasmStateVal)
    (ρCasmStateVal : CasmStateVal)
    (num_steps: Nat) : Prop :=
  num_steps < 2^29 → casmStateVal.strongly_bounded₀ num_steps →
    ρCasmStateVal.strongly_bounded (num_steps+1)
    ∧ ∀ mem : Felt252 → Felt252,
        memory.Agrees mem →
          ∃ instruction : Instruction,
            mem casmStateVal.pc.toFelt252 = instruction.toNat
            ∧ instruction.NextState mem
                casmStateVal.toRegisterStateFelt252 ρCasmStateVal.toRegisterStateFelt252
\end{lstlisting}
It says that for any memory assignment, assuming the input casm state registers are suitably bounded (see Section~\ref{subsection:casm:address:bounds}), there is an instruction in memory such that the output casm state is the appropriate next state for the instruction according to the virtual machine semantics.

Establishing correctness of the corresponding claim for the Stone encoding is a major part of our initial Stone verification project \cite{avigad:et:al:22}, and, similarly, verifying the correctness of this component was a substantial task here.

\subsection{Multiplication}

The \lstinline{MulOpcode} component implements the multiplication instruction of the assembly language.
To the verifier, the specification of the component is essentially the same as that of the generic opcode. The main difference is that the implementation is limited to multiplication instructions. (We also verify that when the constraints are satisfied, the instruction is, in fact, multiplication, but that information is not needed for the soundness proof.) The bulk of the work is carried out by a procedure known as \lstinline{VerifyMul252}.

For Felt252 operands $x$, $y$, and $z$ determined by the instruction flags, the mul instruction asserts $z = x \cdot y$ modulo the Felt252 prime, which we now denote $P$. (Recall that $P = 2^{251} + 17 \cdot 2^{192} + 1$.) Thus, \lstinline{VerifyMul252} has to add constraints to the Cairo-AIR that guarantee this is the case. Recall from Section~\ref{subsection:encoding:felt252s} that $x$, $y$, and $z$ are each encoded as 28 M31 felts, each of which, at this stage, has been range-checked to represent 9-bit numbers.

\newcommand{\mi}[1]{\mathit{#1}}

Assuming $x$ and $y$ are represented by the 28-tuple of 9-bit numbers $(X_0, \ldots, X_{27})$ and
$(Y_0, \ldots, Y_{27})$, their values as Felt252s are $\sum_{i < 28} X_i 2^{9i}$ and $\sum_{i < 28} Y_i 2^{9i}$. Their product is therefore the Felt252 represented by
\[
\Big(\sum_{i < 28} X_i 2^{9i}\Big) \cdot \Big(\sum_{i < 28} Y_i 2^{9i}\Big) =
\sum_{i, j < 28} X_i Y_j 2^{9 (i + j)} = \sum_{d < 55} \Big( \sum_{i < 28} X_i Y_{d - i}\Big) 2^{9 d}
\]
modulo $P$. The right-hand side is the familiar convolution sum that groups together the terms with a common exponent. For each value of $d$, the inner sum is no longer a 9-bit number, but easily small enough to fit inside an M31 Felt, so the product of two Felt252s can naturally be expressed as a tuple of 55 Felts. This provides one strategy for the prover to establish the desired claim:
\begin{itemize}
\item Let $(W_0, \ldots, W_{54})$ be the 55 values of the convolution: $W_d = \sum_{i < 28} X_i Y_{d - i}$, and let $w$ be the product of $x$ and $y$ as integers. Note that $W$ provides a base $2^9$ representation of $w$, although individual ``digits'' are not required to be less than $2^9$.
\item Let $(Z_0, \ldots, Z_{27})$ be the base $2^9$ digits of $z = x \cdot y \mod{P}$. Note that $Z$ and $W$ differ in two ways: the digits of $W$ may exceed the $2^9$ bound, and $W$ may be larger than $P$. We check that $z \equiv w \pmod{P}$.
\end{itemize}

For both steps, note that the verification follows the common pattern described in Section~\ref{subsection:encoding:felt252s}: constraints are added that ensure equations between expressions involving Felts, and part of the verification task is to write down the corresponding natural-number expressions and show that they are bounded by the M31 prime, so that the identities transfer to the latter representations.

The S-two prover uses careful optimizations for both parts. Note that verifying that the initial 55 digits representing the product are calculated correctly requires several constraints over sums and products involving the digits of $x$ and $y$. A clever optimization, namely Karatsuba's algorithm, can be used to reduce the number of constraints. Suppose $u$ and $v$ are base $b$ numbers $u = u_1 b + u_0$ and $v = v_1 b + v_0$. Their product is equal to
\[
  u_1 v_1 b^2 + (u_1 v_0 + u_0 v_1) b + u_0 v_0.
\]
Let $c_0 = u_0 v_0$, $c_1 = u_1 v_0 + u_0 v_1$, and $c_2 = u_1 v_1$ be the coefficients in the expansion above.
Calculating $c_0, c_1, c_2$ seems to require four multiplications, but Karatsuba's trick saves one of them: note that the middle coefficient $c_1$ is $(u_1 + u_0) (v_1 + v_0) - c_2 - c_0$, requiring a total of only three multiplications.

If we split the two 28-digit base $2^9$ numbers $x$ and $y$ into two parts,
\[
  x = x_1 \cdot 2^{14 \cdot 9} + x_0 \quad \text{and} \quad y = y_1 \cdot 2^{14 \cdot 9} + y_0,
\]
the multiplication of 28-digit numbers reduces to three multiplications of 14-digit numbers. Applying the trick again, the latter can be expressed in terms of multiplications of 7-digit numbers, which are calculated with the usual convolution identity.

This takes us to the second step: the prover needs to establish that $w - z \equiv 0 \pmod{P}$. Recall that the digits of $W$ may exceed the $2^9$ bound, but they are still manageable. Let $U_i = W_i - Z_i$ be the pointwise differences of the digits,
and let $u = \sum_{i < 55} U_i 2^{9i}$ be the corresponding integer. We need to add constraints to ensure that $u \equiv 0 \pmod{P}$.

Notice that $u$ will generally be much larger than $P$, in fact, on the order of $P^2$. The strategy is to carefully reduce its value modulo $P$. Actually, we do something even more devious.
Rather than showing that $u \equiv 0 \pmod{P}$, we show $u \cdot (2^{64} + 17 \cdot 2^{5}) \equiv 0 \pmod{P}$. As $2^{64} + 17 \cdot 2^{5}$ is invertible modulo $P$, the two claims are equivalent.

Split $u$ as follows:
\[
u = 2^{49 \cdot 9} u_2 + 2^{21 \cdot 9} \cdot u_1 + u_0,
\]
where $u_0$ corresponds to the 21 least significant digits in $U$, $u_1$ corresponds to the next 28 digits, and $u_2$ corresponds to the 6 most significant digits.
Calculating, we have
\begin{align*}
  u \cdot (2^{64} + 17 \cdot 2^{5}) &=
2^{49 \cdot 9} (2^{64} + 17 \cdot 2^{5}) u_2 +
2^{21 \cdot 9} (2^{64} + 17 \cdot 2^{5}) \cdot u_1 +
(2^{64} + 17 \cdot 2^{5}) u_0
\end{align*}
and
\begin{align*}
  2^{49 \cdot 9} \cdot (2^{64} + 17 \cdot 2^{5}) & \equiv 2^3 + 17 \cdot 2^{195} \\
  2^{21 \cdot 9} \cdot (2^{64} + 17 \cdot 2^{5}) & \equiv -4
\end{align*}
modulo $P$, so it suffices to add constraints that guarantee that
\[
  (2^3 + 17 \cdot 2^{195}) u_2 - 4 u_1 + (2^{64} + 17 \cdot 2^5) u_0 \equiv 0 \pmod{P}.
\]
Call this expression $\widehat u$. These representations are now small enough that the prover can choose a sufficiently small value of $k$ and show that $\widehat u - k \cdot P = 0$. More precisely, one can calculate the non-normalized digits of that expression pointwise; they may be positive or negative, and outside the range $0 \ldots 2^9$, but they still have moderate bounds, and normalizing them to base $2^9$ yields $0$. The component adds exactly these constraints, namely, those that guarantee that normalizing the digits of the calculation yields 0. As always, these constraints are expressed on the Felt252 values; the verification involves showing that the corresponding integer representations are small enough to guarantee that the equations hold there, as well as formalizing the reduction above.

\subsection{Casm address bounds}
\label{subsection:casm:address:bounds}

Variables in Cairo programs take values in the Felt252 finite field, but they are generally intended to represent integers. Range checks serve to guarantee that they represent integers within a certain range, and proving the correctness of a Cairo program generally requires bounding calculations to ensure that they don't wrap around the finite field to yield misleading results. As described in \cite{avigad:et:al:25}, sometimes the relevant bounds can depend on the number of steps that have been executed by the virtual machine. For example, an iterative procedure that adds one to a variable until some condition is met will overflow if the loop continues too long, and it is often necessary to use the fact that the total number of execution steps is bounded.

The conclusion of the soundness theorem described in Section~\ref{section:soundness} requires a bound on the number of steps, for another reason: in the Cairo semantics, memory is modeled as a function from the Felt252 field to itself, and the three registers---the program counter (pc), allocation pointer (ap), and frame pointer (fp)---can be arbitrary Felt252s. This makes sense, since all the registers serve as pointers to memory.

However, when designing the S-two encoding, StarkWare decided that for the types of programs the company wants to certify, it was safe to assume that each of the registers would always fit inside a single M31 Felt, and thus they are represented as such in the encoding. This is a completeness issue rather than a soundness issue. The conclusion of the soundness theorem in Section~\ref{section:soundness} asserts the existence of an execution trace with respect to the original Felt252 semantics. The hypotheses assume that the registers in the initial state are bounded by $2^{29}$ (recall that the M31 prime number is $2^{31}-1$) and that the program runs for at most $2^{29}$ steps.

Even with these constraints, proving soundness is delicate. The soundness theorem asserts the existence of an assignment to Felt252 memory consistent with the publicly shared data, and one can fill memory locations that can never be accessed with any arbitrary value, such as 0. But the soundness theorem also requires that calculations with the registers are consistent with their interpretation in the Felt252 virtual machine semantics. Because at times one may subtract an offset from a memory address, it is useful to take the M31 Felt to encode a range that includes negative numbers as well. Specifically, we chose the range $[-(2^{29} + 1), 2^{31} - 2^{29})$, so that a Felt is interpreted as a Felt252 according to this definition:
\begin{lstlisting}
  def Felt.toFelt252 (x : Felt) : Felt252 := (x + 2^29 + 1).val - (2^29 + 1)
\end{lstlisting}
Read this as follows: given a Felt $x$, to interpret it as a Felt252, add $2^{29} + 1$, take the principal value modulo $2^{31} - 1$ in the range $[0, 2^{31} - 1)$, cast that to a Felt252, and then subtract $2^{29} + 1$. This has the net effect that Felts represented by numbers at the high end of the range $[0, 2^{31} -1)$ are interpreted as negative integers and represented accordingly as Felt252s. This often requires fiddly calculations to show that the cast from a Felt to a Felt252 respects arithmetic operations. For example:
\begin{lstlisting}
theorem toFelt252_add {n a : Nat} (h : n + a < Stwo.P - 2^29 - 1) :
    ((↑n : Felt) + ↑a).toFelt252 = (↑n : Felt).toFelt252 + ↑a
\end{lstlisting}
In words: adding an offset \lstinline{a} to a Felt representing \lstinline{n} commutes with casting \lstinline{n} to a Felt252 as long as the principal value of \lstinline{n + a} is not part of the high-end of the range that is interpreted as negative.

During the execution of a Cairo program, the allocation pointer ap can increase by 1, by 2 (in the call opcode), or by an arbitrary value (in the increase-ap opcode). The frame pointer fp can either change to the current value of ap or change back to a previous value.
Our soundness proof shows that, assuming the registers are initially bounded by $2^{29}$, an invariant of the execution trace is that at each step, the registers are bounded by $2^{29}$ plus the number of steps.
\begin{lstlisting}
def strongly_bounded (s : CasmStateVal) (num_steps : Nat) : Prop :=
  (∃ n : Nat, (n < 2^29 + num_steps) ∧ s.ap = ↑n) ∧
    (∃ m : Nat, (m < 2^29 + num_steps) ∧ s.fp = ↑m)
\end{lstlisting}

The dependence of the bound on the number of steps handles the case where \lstinline{ap} increases by 1. For the increase-ap opcode, there are designated constraints that bound \lstinline{ap} by $2^{29}$.
For the call opcode, we use the fact that it accesses the memory at \lstinline{ap}, and the Cairo-AIR encoding checks that accessed addresses are bounded. To handle a small corner case, we have to assume a slightly stronger bound on the allocation pointer when the number of steps is zero, though we can only ensure the weaker bound after that. That explains the slight variant of the previous definition in some hypotheses:
\begin{lstlisting}
def strongly_bounded₀ (s : CasmStateVal) (num_steps : Nat) : Prop :=
  (∃ n : Nat, (n < 2^29 - 1 + if num_steps = 0 then 0 else 1 + num_steps) ∧ s.ap = ↑n) ∧ (∃ m : Nat, (m < 2^29 + num_steps) ∧ s.fp = ↑m)
\end{lstlisting}

An early implementation, before S-two was used in production, omitted the additional check needed to maintain the invariant in the case of the increase-ap opcode. Our verification uncovered this oversight, and the Cairo-AIR encoding was fixed accordingly.

\section{Logups}
\label{section:logups}

The logup constraints provide a communication system between the components. They establish either a subset relation or an equal-counts relation between the use tuples and the yield tuples of each \emph{relation}. As these tuples are registered by different components, this transfers properties established by one component to another. It can also establish the existence of valid sequences (for example, of states) where each step in the sequence comes from a different component. The construction of the logup constraints involves several steps, where the tuples are first mapped to single field elements and then constraints are defined on these elements. We will present this construction backward, starting with the constraints on the single field elements and then describing the mapping from the tuples.

\subsection{The logup lemma}
\label{subsection:the:logup:lemma}

At the core of the logup mechanism is the following lemma, which is a formalization of (one direction of) lemma 5 in \cite{habock:22}:

\begin{lstlisting}
lemma count_eq_of_sum_eq' {na nb : Type _} [Fintype na] [Fintype nb]
    (a : na → F)
    (b : nb → F) :
  ∀ m : nb → F,
    ∑ i : na, RatFunc.mk 1 (X - C (a i)) =
        ∑ i : nb, RatFunc.mk (C (m i)) (X - C (b i)) →
      ∀ z, ↑(Multiset.count z (Multiset.map a Finset.univ.val)) =
          μ (fun i => b i) (fun i => m i) z
\end{lstlisting}

This lemma takes two multisets $a$ and $b$ of values in a finite field $F$ (represented as functions from an index type to a value), and considers two sums of rational functions based on these values, $\sum_{i}\frac{1}{X - a(i)}$ and $\sum_{i}\frac{m(i)}{X - b(i)}$. The notation \lstinline{C (a i)} converts the value \lstinline{a i} to the corresponding constant polynomial, and similarly for \lstinline{C (m i)} and \lstinline{C (b i)}. The lemma asserts that whenever one can find multiplicities $m(i)$ that make these two sums equal (as functions), we know that for each value $z$, the number of times $z$ appears in $a$ is equal to the sum of the multiplicities $m(i)$ over the indices $i$ such that $b(i) = z$. For the multiset $a$, each value appears with multiplicity 1, while for $b$, we may choose the multiplicities $m$. The Lean expression \lstinline{μ (fun i => b i) (fun i => m i) z} denotes the sum of $m(i)$ over the indices $i$ such that $b(i) = z$. Note that the equality stated in the lemma is between field elements, not integers. To interpret these sums as counts, we have to be careful and make sure the sum does not exceed the characteristic of the field. We were therefore required to introduce proper assumptions on set sizes in the relevant lemmas.

The logup mechanism constructs polynomial constraints which (with high probability) imply that the sums of rational functions in this lemma are equal, thus establishing the premise of the lemma. The values $a$ are the use values and the values $b$ are the yield values. We can then use this basic lemma in two different ways. When the multiplicities $m$ are not agreed on in advance, the fact that such multiplicities exist implies that every value in $a$ is also a value in $b$. Such a subset relation is needed when we want to know that a certain value has some required property. The yield values (the $b$) are known to have the property (for example, they are range checked) and the satisfaction of the logup constraints implies that also the use values (the $a$), as a subset of the yield values, have this property. The same value can be ``used'' many times (i.e.,~can appear multiple times in the multiset $a$) against a single entry of the same value in the set of yields. We refer to this use of the logups as a \emph{subset relation}.

Another way to use this lemma is to require that all the multiplicities $m$ assigned to the yield values are 1. From the lemma we can then conclude that each value appears in the use multiset $a$ exactly the same number of times as in the yield multiset $b$. We refer to this use of the logups as a \emph{chain relation}. The chain relation is used when we have a set of pairs, $(u,y)$, generated by different components, where $u$ is a use value and $y$ is a yield value. If we view these pairs as directed edges in a graph, the equal-count property at each vertex implies that the edges form a cycle. When the pairs are the input and output states of opcode components together with a pair from the final state to the initial state, this shows that there is a sequence of state changes starting with the initial state and ending in the final state.

\subsection{Polynomial constraints}
\label{subsection:polynomial:constraints}

In the Cairo-AIR, the values $a$ and $b$ in the logup lemma come from calls to the components, as described in Section~\ref{subsection:components:and:messages}.
To show that the premise of the logup lemma holds, the AIR encodes the sum of rational functions as a sequence of partial sums (according to components and rows), with an additional variable tracking the cumulative sum. First, the equality of sums $\sum_{i}\frac{1}{X - a(i)} = \sum_{i}\frac{m(i)}{X - b(i)}$ is replaced by $\sum_{i}\frac{1}{X - a(i)} - \sum_{i}\frac{m(i)}{X - b(i)} = 0$. Next, we merge the use and yield values and partition the sum into partial sums: $\sum_k \sum_{\{i | s(i) = k\}}\frac{m(i)}{X - f(i)}$, where the values $f$ are all the values in $a$ and $b$ together (in some arbitrary order), the multiplicities $m$ are 1 for values in $a$ and are the negation of the $b$ multiplicities, and $s(i)$ defines the partial sum that index $i$ belongs to.

\newcommand{\psum}{\mathit{psum}}  
\newcommand{\pr}{\mathit{pr}}

The terms in the sum will be further grouped into one or more \emph{partitions}, indexed by a function $\pr$ from the indices of $f$ to the index of the partition to which $i$ belongs. The partitioning defined by $\pr$ is independent of the partitioning defined by $s$. Unlike the $s$-partitioning, which is dictated by the AIR construction and may freely mix terms from different relations, the $\pr$-partitioning respects relations: each relation is contained in a single partition, although a single partition may contain several relations whose values are guaranteed to be disjoint. The motivation for the $\pr$-partitioning will become clear in the next section, where we discuss multiple relations.

We evaluate each term in the sum on a value $z(\pr(i))$. This results in the equations:
\[
\sum_{\{i | s(i) = k\}}\frac{m(i)}{z(pr(i)) - f(i)} = psum(k+1) - psum(k)
\]
where $\psum(k+1)$ is the cumulative partial sum up to $k$, beginning with some arbitrary $\psum(0)$. Assume that the values of $s$ are $0, 1, \ldots, n_s$.
It follows from $\psum(n_s + 1) = \psum(0)$, and from these partial equations, that the full sum of evaluated rational functions is equal to zero. These equations are converted into equivalent polynomial equations:
\[
\sum_{\{j | s(j) = k\}} m(j) \prod_{\{i | s(i) = k \land i \ne j\}} (z(pr(i)) - f(i)) = (psum(k + 1) - psum(k)) \prod_{\{i | s(i) = k\}} (z(pr(i)) - f(i))
\]
The equations are defined in the Lean code:
\begin{lstlisting}
def cumulativeC_k {t n_s n_p : Nat} (f m : Fin t → F) (s : Fin t → Fin n_s)
      (pr : Fin t → Fin n_p) (psum : Nat → F) (z : Fin n_p → F) (k : Nat) :=
  let f_k := funK s k f
  let z_k := funK s k (z ∘ pr)
  let m_k := funK s k m
  ∑ j : Fin (indxsK s k).length, (m_k j) * (prodZ_exc f_k z_k j) =
      (prodZ f_k z_k) * (psum k.succ - psum k)
\end{lstlisting}
where \lstinline{indxsK s k} are the indices of the $k$th partial sum, as defined by $s$, \lstinline{funK} restricts a function to this set of indices, and \lstinline{prodZ} and \lstinline{prodZ_exc} are the full product and the product excluding $j$.

The AIR defines multivariate polynomials that, when evaluated under the component assignments, result in the equations defined here. The values $z$ are chosen using the Fiat–Shamir heuristic, and function as random values. With high probability, the satisfaction of the equations with these $z$ values implies the premise of the logup lemma.

Note that there are many ways to split the sum into partial sums (the partition defined by $s$), and the split used by the AIR depends on the exact way in which the AIR components add their polynomial constraints.

\subsection{Relations and partitions}

The logup lemma provides a way to establish a property (subset relation or equal counts) between two multisets. In the Cairo-AIR we would like to establish such a property for multiple pairs of multisets, representing different relations. For example, the range check is one relation, and the memory is defined by two relations. The AIR encodes the logups for all these relations in a single cumulative sum. There are two ways to ensure that the required properties hold separately for each relation. The first is to ensure that values in different relations can never be the same, for example, by using the first coordinate of the tuple to represent the relation identifier. The second is to use multiple values of $z$. The indices in the sum for which a single value of $z$ is used are called a partition. The logup lemma then holds (with high probability) for each partition separately. The two methods can be combined, with multiple partitions and with several relations with disjoint values in a single partition (in the process of developing the AIR, the actual approach used changed from a single relation per partition to a single partition for all relations).




The next lemma provides the conditions under which the subset property can be inferred for a single relation:
\begin{lstlisting}
lemma inclusion_of_constraints_and_not_exceptionalSet' ...
\end{lstlisting}
The lemma is given a set of indices \lstinline{use_i} which are the indices $i$ for which $f(i)$ is a use value of the relation the lemma is applied to. These indices are assumed to satisfy two properties. First, that the multiplicities assigned to them by $m$ are all 1, and, second, that the number of indices is bounded by the characteristic of the finite field. This ensures that the sum on their multiplicities (which are in the finite field) cannot be zero. The lemma assumes that all the cumulative constraints \lstinline{cumulativeC_k} (see~\ref{subsection:polynomial:constraints}) are satisfied for $k = 0, \ldots, n_s$ and that $\psum(n_s + 1) = \psum(0)$. It then shows that if the $z$ which appears in these equations is not in some set to be avoided (the \emph{exceptional set}) then the values of $f$ on the set of indices \lstinline{use_i} are a subset of the values of $f$ on all other indices in the same partition. Later, when we add the assumption that values from different relations in the same partition do not collide, this will mean that the use values are a subset of the yield values of the same relation.

The interesting part in this lemma is the choice of $z$. There is a set of values, the exceptional set, for which the conclusion of the lemma does not necessarily hold. This set, \lstinline{exceptionalSet f m pr}, depends on the values $f$ and $m$ and the partitions, but not on the relations. It consists of the values of $z$ that make the cumulative sum equal zero even when the sums of poles that appear in the logup lemma are not equal. For any other choice of $z$, the premise of the logup lemma holds for each relation separately. As the value $z$ can be seen as a random choice, it is enough to show that the exceptional set is small. The following lemma provides a bound on the size of the exceptional set:
\begin{lstlisting}
lemma card_exceptionalSet_le {t n_p : Nat} (f m : Fin t → F)
    (pr : Fin t → Fin (n_p + 1)) :
  (exceptionalSet f m pr).card ≤
    Fintype.card F ^ n_p * t + Fintype.card F ^ n_p * max_pr_len pr
\end{lstlisting}
The number of possible choices of $z$ is $|F|^{n_p + 1}$, so the probability of a bad choice is no larger than $\frac{t + l}{|F|}$ where $t$ is the number of values in all logup constraints, $l$ is the maximal number of values in a single partition, and $|F|$ is the size of the finite field. While the tuples we encode in the logups are tuples of \lstinline{Felt}s, the logup constraints use the degree 4 extension field of the \lstinline{Felt} field. This increases $|F|$, decreasing the probability of $z$ falling into the exceptional set.

A very similar lemma holds for chain relations, for which we need to establish that, for each value $x$, the value's count in the use multiset is the same as its count in the yield multiset.
\begin{lstlisting}
lemma equal_count_of_multiplicity_one'' ...
\end{lstlisting}
This lemma is very similar to the previous lemma. The crucial difference is the added assumption that the multiplicities of the yield values are -1 (which is equivalent to a multiplicity of 1 in the logup lemma). We must also introduce the assumption that the values of different relations in the same partition are disjoint. This assumption is also needed for the subset relations, but for those it is introduced at a later stage.

\subsection{Relations on tuples}

The lemmas we discussed so far are about single field elements. For most actual uses, however, we need the subset and equal-count properties to hold for \emph{tuples}. For example, the memory is defined by two relations, one mapping addresses to IDs, and one mapping IDs to values. The address to ID mapping is defined as a set of (address, ID) pairs, while the ID to value mapping is defined by a tuple of size 29: one Felt for the ID and 28 Felts to describe the Felt252 value. To apply the logup lemma to tuples, we map the tuples to single field elements using the polynomial function $combine(a) = \sum_{i=0}^{n-1} a(i) \alpha^i$:
\begin{lstlisting}
def combine {n : Nat} (α : F) (a : Fin (n + 1) → F) :=
  ∑ i : Fin (n + 1), a i * α ^ (i : Nat)
\end{lstlisting}
The value $\alpha$ is chosen using the Fiat–Shamir heuristic (similarly to the $z$'s in the previous lemmas) and functions as a random value. For each partition we can choose a different $\alpha$, as each partition provides a separate application of the logup lemma.

As long as the \lstinline{combine} function does not map two different tuples to the same field element, the previous lemmas, applied to the combined values, would immediately imply that the same properties (subset or equal counts) also hold for the tuples. However, there is no guarantee that the \lstinline{combine} function does not map different tuples to the same field element. We must therefore define a \emph{bad set} for the choice of $\alpha$, similar to the bad set of choices for $z$. Only a choice of $\alpha$ outside this set guarantees the conclusion we want, but if this bad set is sufficiently small, the probability of failure is also sufficiently small.

If we were to require a choice of $\alpha$ such that no two different tuples (in the sets of use and yield tuples) collide, the bad set would become too large for our purposes, in fact, quadratic in the size of the partition (without any further assumptions on the distribution, the number of possible values of $\alpha$ is only bounded by the number of pairs of tuples, which is quadratic in the number of tuples). However, for the desired conclusions to hold, we can use a more restricted definition of the bad set. Since the bad set depends on the use and yield values, we define the bad set to be the set of values of $\alpha$ that would cause us to reach an incorrect conclusion about the use and yield values. In other words, given use and yield values which do not satisfy the required property (such as the subset property) the bad set is the set of $\alpha$'s for which the constraints would still hold in this case (and would lead us to conclude that the subset relation holds even though it does not). As a result, we need two distinct notions of badness, one for the subset relations and one for the chain relations. The bad set for the subset relations is defined as follows:
\begin{lstlisting}
def badSet {n l: Nat} (tuples : Fin l → (Fin (n + 1) → F)) (use_i : Finset (Fin l)) (yield_i : Finset (Fin l)) : Finset F :=
  if Finset.image tuples use_i ⊆ Finset.image tuples yield_i
  then ∅
  else
    Finset.univ.filter
      fun α => ∀ u, u ∈ Finset.image tuples use_i \ Finset.image tuples yield_i →
        ∃ y, y ∈ Finset.image tuples (Finset.univ \ use_i) ∧ combine α u = combine α y
\end{lstlisting}
When the use values are a subset of the yield values, the bad set is empty; the property we want holds, so we do not care what $\alpha$ is. Otherwise, if there are some use tuples which are not yield tuples for the same relation, the bad set consists of those values $\alpha$ such that for every use tuple $u$ that is not a yield tuple for the same relation, the \lstinline{combine α} function maps the use tuple $u$ to the same value as some other tuple in the partition. Such an $\alpha$ hides the fact that the use tuple is not equal to some yield tuple (in the same relation). An $\alpha$ is bad only if it hides all such use tuples.

A similar bad set is defined for the chain relations, but instead of using the subset property, the definition uses the equal count property. If the counts are not equal, the bad set consists of all those $\alpha$'s that map use tuples for which the counts are not equal to other yield tuples (thus possibly making the counts seem equal).

In both cases, the size of the bad set is linear in the size of the partition. This is because we only need to exclude $\alpha$'s that hide all violations of the relation, so we can choose one such violation, for a use tuple $u$, and bound the number of $\alpha$'s that hide the violation for $u$ by mapping it to the same field element as some other tuple $y$. The number of such $\alpha$'s is bounded by the number of tuples $y$ (times the size of the tuple - 1, as this solves a polynomial equation). For the subset property this is given by the following lemma:
\begin{lstlisting}
lemma card_badSet_le {n l: Nat}
      (tuples : Fin l → (Fin (n + 1) → F))
      (use_i : Finset (Fin l))
      (yield_i : Finset (Fin l))
      (h_disjoint : ∀ i ∈ use_i ∪ yield_i, ∀ j ∉ use_i ∪ yield_i, tuples i ≠ tuples j) :
    (badSet tuples use_i yield_i).card ≤ l * n
\end{lstlisting}
Here, $l$ is the size of the partition (number of lookup terms, both use and yield, of all relations in the partition). The value $n$ is one smaller than the size of the tuples in the partition (all tuples in a partition are assumed to be of the same size, by padding them with zeros when necessary). As $\alpha$ is chosen randomly from the field $F$, the probability of it being in the bad set is at most $\frac{l\,n}{|F|}$.

Having defined the bad sets, we extend the logup claims from multisets of field elements to multisets of tuples, by defining a mapping \lstinline{to_all_indxs} from the indices of the tuples to the indices in $f$ of their combined field elements. As long as $\alpha$ is not in the bad set, the logup lemmas for the values in $f$, described in the previous section, induce the same property (subset or equal-count) on the tuples.

\subsection{Logup constraint construction}
\label{subsection:logup:constraint:construction}

To reason about tuples, relations, logup constraints, and all the other ingredients that go into the lookups, we define a structure that collects all values involved in the lookups and all assumptions required to ensure that this is a proper construction.
\begin{lstlisting}
structure LookupsSatisfied [Fact (Nat.Prime Stwo.P)] (t n_s n_p n_r : ℕ)
    (tl : Fin (n_p + 1) → Nat)
    (rp : Fin (n_r + 1) → Fin (n_p + 1))
    (chain_rel : Finset (Fin (n_r + 1))) where
  values : LookupValues t n_s n_p
  constraints : LookupConstraints values
  partitions : (p : Fin (n_p + 1)) → LookupPartition values p (tl p)
  tuples : (k : Fin (n_r + 1)) → RelationTuples (partitions (rp k))
  covering: RelationsCoverPartitions tuples
  h_inj: ∀ k ∈ chain_rel,
    (fun (i : {i // i ∈ (tuples k).use_i ∪ (tuples k).yield_i}) =>
      (partitions (rp k)).to_all_indxs i).Injective
  h_chain: ∀ k ∈ chain_rel,
    ∀ i ∈ (tuples k).yield_i, values.m ((partitions (rp k)).to_all_indxs i) = -1
  h_bad : ∀ k, NotInBadSet k chain_rel (tuples k)
\end{lstlisting}
The substructure \lstinline{values} defines the values $f$, the multiplicities $m$, and their partitions. The substructure \lstinline{constraints} defines the cumulative constraints based on the values and partitions in \lstinline{values} and adds the assumption that the constraints are satisfied with a value $z$ that is not in the exceptional set. The substructures \lstinline{partitions} define the tuples in each partition and their mapping to the values of $f$ using the \lstinline{combine} function. The assumption \lstinline{h_bad} tells us that the $\alpha$ chosen for each partition is not in the bad set of any of the relations. Additional assumptions that this is a proper construction are also included (for example, that every value in $f$ is the image of some tuple). Finally, the substructures \lstinline{tuples} define the subset of indices in each partition which belong to the use and yield values of each relation. Special assumptions which are only needed for the chain relations (defined by the subset \lstinline{chain_rel}) are added by \lstinline{h_inj} (no two tuples are mapped to the same element of $f$, as this will distort counting) and \lstinline{h_chain} (the multiplicities of yield values are -1).

Together, these define any valid construction of relation tuples, their mapping to field elements, and the cumulative constraints on these values, together with the assumption that the constraints are satisfied and that the choices of $z$ and $\alpha$ do not fall into the bad sets. One important detail is that the values $f$, $m$, $z$, and $\alpha$ are all in the field \lstinline{LookupF} (the degree 4 extension of the field \lstinline{Felt}) while the tuples are tuples of \lstinline{Felt}s. The lifting takes place when applying the \lstinline{combine} function. Choosing $z$ and $\alpha$ out of a larger field decreases the chance that they fall into the bad sets.

The \lstinline{LookupsSatisfied} structure contains all the assumptions needed for the subset or chain property to hold, except for one assumption that depends on the tuples themselves: that tuples of different relations in the same partition are never the same. As this is a property of the tuples, it must be deduced from the construction of the tuples by the components.

\subsection{Component lookups}
\label{subsection:component:lookups}

Up to this point, we described arbitrary multisets of tuples. The next step is to apply the lemmas to the tuples generated by the components. In Section~\ref{subsection:components} we described how the lookup terms for each component are evaluated on the assignments to that component, and all the resulting tuples are collected into a single long list, \lstinline{ComponentLookups}. To these, we add the external lookups, which include the public memory, the initial state, and the final state, to produce the list of all lookups, \lstinline{AllLookups}.

This long list of lookups can be split by relation and by whether they are use or yield lookups. Once they are split, we can discard the extra information in the structures and only keep the tuples (as only the tuples appear in the logup constraints).
All that remains to do is assume a valid \lstinline{LookupsSatisfied} construction where the tuples (for each relation and use/yield choice) are exactly the tuples defined by the components. These are exactly the assumptions that appear in the final soundness theorem (see Section~\ref{subsection:soundness}):
\begin{lstlisting}
  -- Lookups are satisfied
  (h_satisfied : LookupsSatisfied t n_s NUM_LOOKUP_REL_MINUS_ONE rel_lengths chain_rels)
  -- Lookups agree with the assignments to the lookup terms
  (h_use_agree : ∀ rel, RelUseLookupsAgree inp pubMemLookups varAssigns rel (h_satisfied.tuples rel))
  (h_yield_agree : ∀ rel, RelYieldLookupsAgree inp pubMemLookups varAssigns rel (h_satisfied.tuples rel))
\end{lstlisting}
The assumption \lstinline{h_use_agree} says that for every relation, the use tuples created for the input \lstinline{inp}, the public memory lookups \lstinline{pubMemLookups} and by evaluating the components on the variable assignments \lstinline{varAssigns}, are exactly the same (as a multiset) as the tuples \lstinline{h_satisfied.tuples rel}, which are the tuples defined for this relation in \lstinline{LookupsSatisfied}. A similar assumption must also hold for the yield tuples.

\section{Conclusion}
\label{section:conclusions}

We have described a formal verification, in the Lean 4 proof assistant, of the Cairo-AIR encoding. Specifically, we have modeled, in Lean, the code generating the constraints and lookups corresponding to the AIR components, as well as the logup constraints that guarantee that, with high probability, the protocol for passing messages between components is sound. Our soundness theorem states that assuming the components are put together correctly and that the logup constraints are met, the AIR encoding ensures the existence of a Cairo execution trace that is consistent with data that is publicly shared between the prover and the verifier. The proof relies on the same formal specification of the Cairo virtual machine semantics that we used to verify the Stone AIR encoding \cite{avigad:et:al:22}.

Our soundness theorem requires trust that the S-two code has been modeled in Lean correctly, and that the ambient hypotheses, describing some of the ambient S-two bookkeeping, are correct. Bracketing these concerns allowed us to focus on the most delicate parts of the encoding, where soundness assurances were most valuable. During the design of S-two, our verification turned up an instance where additional range checks were needed to ensure soundness, and insufficient security bounds on the logup protocol convinced StarkWare's engineers to modify the design to increase the number of security bits. Perhaps most importantly, StarkWare's engineers could feel more comfortable pursuing aggressive optimizations, knowing that these optimizations would be formally verified.

\bibliographystyle{plain}
\bibliography{stwo}

\appendix

\section{The formal soundness theorem}
\label{section:formal:statement}

The following is the full statement of our main soundness theorem.
\begin{lstlisting}[xleftmargin=0em]
theorem trace_sound [Fact (Nat.Prime Stwo.P)] [Fact (Nat.Prime Felt252Prime)] {t n_s : ℕ}

      (inp : InputData)
      (pubMemLookups : PublicMem.Lookups)
      (varAssigns : Component → Array VarAssign)

      (h_pubMem : PublicMem.LookupsAgree inp.mStar pubMemLookups)

      -- Lookups are satisfied
      (h_satisfied: LookupsSatisfied t n_s NUM_PARTITIONS NUM_LOOKUP_REL_MINUS_ONE
        partition_lengths rel_partition chain_rels)
      -- Lookups agree with the assignments to the lookup terms
      (h_use_agree : ∀ rel, RelUseLookupsAgree inp pubMemLookups varAssigns rel
        (h_satisfied.tuples rel))
      (h_yield_agree : ∀ rel, RelYieldLookupsAgree inp pubMemLookups varAssigns rel
        (h_satisfied.tuples rel))
      -- Memory assignments are functions.
      (h_is_mem_addr_to_id : IsMemAssign (varAssigns .MemoryAddrToId))
      (h_is_mem_id_to_value : IsMemAssign (varAssigns .MemoryIdToValue))
      -- Range check rows are range checked.
      (h_rc : IsRangeCheckAssign (varAssigns .RangeCheck))

      -- The constraints of the components are satisfied.
      (h_components: ComponentsSatisfied varAssigns)

      -- Number of steps
      (h_num_steps :
        (RelLookups inp pubMemLookups varAssigns OPCODE_TRACE_REL_INDEX .use).size ≤ 2^29)
      (h_state_bound : (initialState inp).strongly_bounded₀ 0) :

    ∃ mem : Felt252 → Felt252,
      Option.FnExtends mem inp.mStar ∧
      ∃ n : Nat, n ≤ 2^29
        ∧ ∃ exec : Fin (n + 1) → RegisterState Felt252,
            exec 0 = (initialState inp).toRegisterStateFelt252
            ∧ exec (Fin.last n) = (finalState inp).toRegisterStateFelt252
            ∧ ∀ i : Fin n, NextState mem (exec i.castSucc) (exec i.succ)
\end{lstlisting}

\end{document}